\documentclass[sigconf]{acmart}

\usepackage[english]{babel}
\usepackage{blindtext}

\usepackage{lipsum}
\usepackage{booktabs}
\usepackage{paralist}
\PassOptionsToPackage{hyphens}{url}
\usepackage{hyperref}
\hypersetup{colorlinks, citecolor=blue, filecolor=blue, linkcolor=blue, urlcolor=blue}
\usepackage{multirow}
\usepackage{subcaption}
\usepackage{graphicx}
\usepackage{xspace}

\usepackage[inline]{enumitem}

\newcommand{\afblock}[1]{\noindent{\textbf{#1}}}

\newcommand{\cwa}[0]{\texttt{CWA}\xspace}

\acmYear{2020}\copyrightyear{2020}
\setcopyright{rightsretained}
\acmConference[SIGCOMM '20 Demos and Posters]{ACM Special Interest Group on Data Communication}{August 10--14, 2020}{Virtual Event, USA}
\acmBooktitle{ACM Special Interest Group on Data Communication (SIGCOMM '20 Demos and Posters), August 10--14, 2020, Virtual Event, USA}
\acmPrice{15.00}
\acmDOI{10.1145/3405837.3411378}
\acmISBN{978-1-4503-8048-5/20/08}

\usepackage{xprintlen}

\begin{document}

\def\UrlBreaks{\do\/\do-}

\begin{abstract}
On June 16, 2020, Germany launched an open-source smartphone contact tracing app (``Corona-Warn-App'') to help tracing SARS-CoV-2 (coronavirus) infection chains.
It uses a decentralized, privacy-preserving design based on the Exposure Notification APIs in which a centralized server is only used to distribute a list of keys of SARS-CoV-2 infected users that is fetched by the app once per day.
Its success, however, depends on its adoption.
In this poster, we characterize the early adoption of the app using Netflow traces captured directly at its hosting infrastructure. 
We show that the app generated traffic from allover Germany---already on the first day.
We further observe that local COVID-19 outbreaks do not result in noticeable traffic increases.
\end{abstract}

\title{Corona-Warn-App: Tracing the Start of the Official COVID-19 Exposure Notification App for Germany}

\author{Jens Helge Reelfs}
\affiliation{%
  \institution{Brandenburg University of Technology}}

\author{Oliver Hohlfeld}
\affiliation{%
  \institution{Brandenburg University of Technology}}

\author{Ingmar Poese}
\affiliation{%
	\institution{BENOCS}}



\maketitle

\section{Introduction: Corona-Warn-App}

The Corona-Warn-App~\cite{coronaApp} (\cwa) is Germany's official contract tracing smartphone app released on June 16, 2020.
It aims to trace infection chains by informing users that were exposed to a person later tested positive.
Centralized contact \emph{tracking} by apps that report contacts to a central infrastructure raise privacy concerns, which is why a decentralized and privacy-preserving contract \emph{tracing} approach (DP-3T) has been proposed~\cite{troncoso2020decentralized}.
This concept evolved to the Exposure Notification APIs by Apple~\cite{ENApple} and Google~\cite{ENGoogle}, of which security and privacy properties were assessed~\cite{baumgrtner2020mind}.
The \cwa uses the decentralized Exposure Notification approach to detect the proximity of other \cwa users by collecting pseudonymous identifiers sent via Bluetooth Low Energy, only stored on the phone.
Its source code---including the Android and iOS smartphone apps, the backend server, and documentation---is released via Github~\cite{CoronaWarnAppGithub}. %

\begin{figure}
  \includegraphics{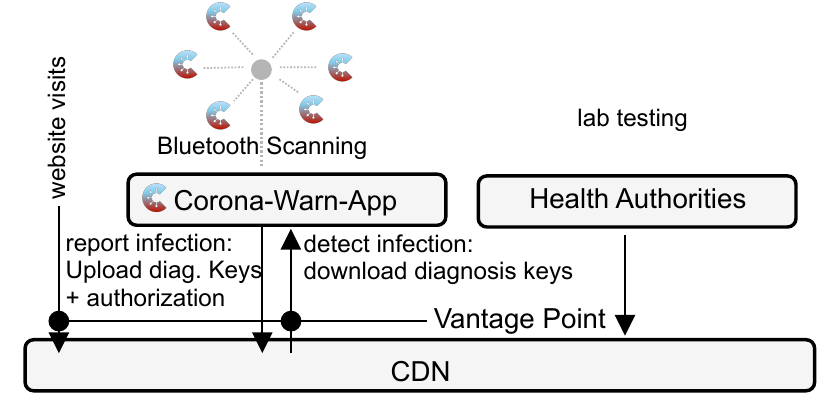}
  \vspace{-2.5em}
  \caption{\cwa architecture and vantage point}
  \label{fig:arch}
  \vspace{-1.8em}
\end{figure}

We show the overall architecture including our vantage point in Figure~\ref{fig:arch}.
Phones locally store these received identifiers for 14 days.
To protect the user's privacy, all identifiers are volatile by generating new temporary exposure keys every 24 hours.
If diagnosed with COVID-19, a user \emph{can} decide to inform others by uploading (parts of her) temporary keys (diagnosis keys) used within 14 days to a central server, verified by health authorities.
By monitoring the API, we observe the first diagnosis keys to be available on June\,23~\cite{cwa-monitor}.
The \cwa regularly downloads shared diagnosis keys from the central server, matches them against the local Bluetooth encounter history, and informs the user of having been exposed to an infected person within the past 14 days if keys match.
Shared keys are non-personal identifiable and all contact tracing data never leaves the phone.

\afblock{Goal.}
Since widespread adoption is key to the app's success~\cite{Ferrettieabb6936}, we take the rare opportunity to monitor its nation-wide adoption starting at day 1.
We measure \emph{interest} in the \cwa by monitoring \cwa app and website traffic at its hosting infrastructure, enabling us to provide first insights into the adoption across Germany.
We further study whether local COVID-19 outbreaks manifest in higher use.

\section{Data Set}

We obtained sampled Netflow traces from routers connecting the data center hosting the \cwa backend (see CDN in Figure~\ref{fig:arch}).
These flows contain web site visits \emph{and} diagnosis key downloads by the app.
All client IP addresses are prefix-preserving anonymized.
We filter server traffic using 2 IPv4 prefixes mentioned in the \cwa backend documentation~\cite{CwaBackendDoc} and omit IPv6.
We verified their usage by resolving the API and web site DNS names (obtained from the app source code) against 10k open DNS resolvers from public-dns.info.
As both, app and website, use HTTPS only, we restrict the data to encrypted HTTPS (tcp/443) IPv4 flows from the CDN to the user---resulting in $\approx3.3M$ matching flows within June 15--25, 2020.

\afblock{Limitations.}
Website visits and \cwa app API calls are served by the same servers via HTTPS and cannot be differentiated.
The routers Netflow cache eviction settings and sampling result in only observing few packets for most flows, making a flow-size based differentiation infeasible.
While \cwa should periodically download diagnosis keys, energy saving settings prohibit background downloads on some Android and iOS phones, reported on July 24~\cite{DWAppDisabled, CWAUpdateBug} and to be fixed after our study.
Periodic request pattern by \cwa might thus be used in future work for app identification.
Yet, the \cwa API DNS name appeared in the Umbrella Top 1M domains~\cite{toplists} on June 24,\,27, July 8,\,10--11, while the website never appeared---implying \emph{\cwa API calls to be more popular than website visits} in OpenDNS and thus \emph{might} dominate the \#flows.
Flows reveal trends in the interest in \cwa and geolocation of destination routers/prefixes enables to study geographic adoption---the scope of this work.

\afblock{Ethics.}
The Netflow data provides only flow summaries based on the packet header and does not reveal any payload information.
All IP addresses are anonymized; it enables us analyzing aggregates of traffic flows between routers (to identify city-level location information of users) but not individual users.
The flow-level statistics do not enable detecting infected users nor deriving \emph{any} user-related information.
Our analyses provide aggregated perspectives on the general interest in the app without compromising users' privacy.
\section{CWA App: Early Adoption Results}

\begin{figure}[t]
  \vspace{-0.8em}
  \includegraphics[width=\columnwidth]{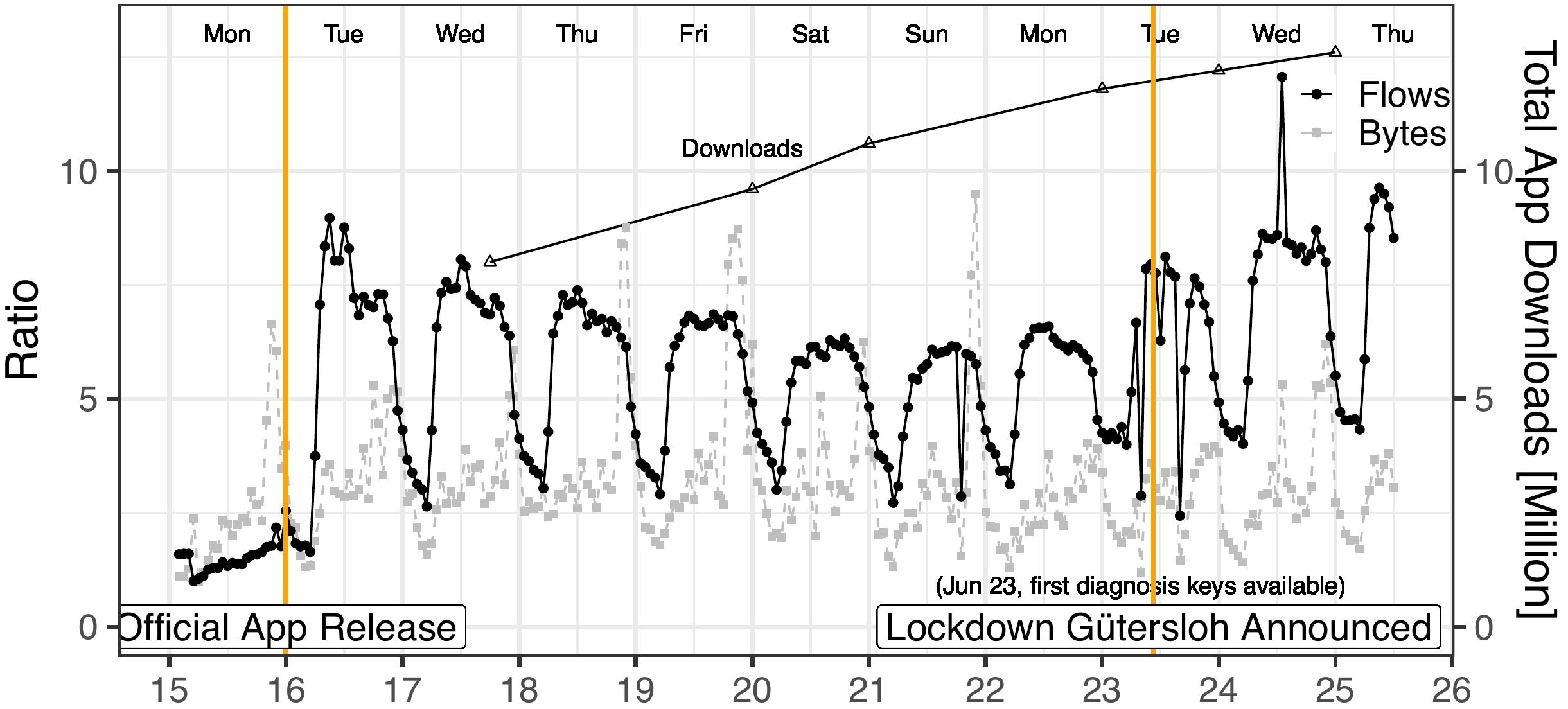}
  \vspace{-2.5em}
  \caption{Hourly aggregated HTTPS traffic from CWA CDN to users normed to the minimum (left y-axis) and the total app downloads in million from Google/Apple (right y-axis).}
  \label{fig:overall}
  \vspace{-1.2em}
\end{figure}

\afblock{Temporal adoption.}
We show all HTTPS traffic \emph{from} the \cwa CDN to its clients in Figure~\ref{fig:overall} (flows and bytes normed to the minimum).
It also cumulative shows officially reported downloads from the Apple and Google playstores~\cite{AppDownloads}, starting on June 17; 36 hours after its release, the \cwa was downloaded 6.4M times (16.2M total downloads by July 24).
With the official release of the \cwa on June~16, the traffic immediately increases ($7.5$x increase of flows on June 16).
Interest starts to follow the normal diurnal traffic pattern. %
After an initial steep traffic increase, it is reduced after a few days, just to re-surge when news in Germany started reporting higher infection rates again and subsequent lockdowns in two districts on June 23~\cite{DWGuetersloh} (G\"utersloh and Warendorf) followed---widely covered in media.

By knowing that customers of certain ISPs keep the same IP address over time, we studied how regular routing prefixes communicate with the \cwa backend (fraction of individual first to last day observed).
We observe sustained interest as 50\% (75\%) of the prefixes occur in 67\% (80\%) of possible days.

\afblock{Quick nation-wide spread.}
The success of the \cwa app to trace infection chains by contact tracing depends on its adoption and geographic spread.
We thus geolocate the request traffic (again both website requests and app API calls---both reflecting interest) within Germany shown in Figure~\ref{fig:map} by ZIP code areas summed over 10 days.
We derive 18\% of geolocations from local routers within an ISP that connect customers (ground truth since the router locations are known), while the rest is located by applying the Maxmind geolocation database on routing prefixes.
Note that client geolocation \emph{can} be subject to errors; the router city-location can be off the clients location (e.g., in rural areas) and Maxmind's geolocation can also be subject to inaccuracies at city-level~\cite{GeolocationIngmar}.
We observe that almost all districts (shown in the heatmap by ZIP code areas) emit requests to the \cwa backend.
Notably, evaluating the geographic spread on the first day of the app leads to almost the same observation (not shown).
In conclusion, the \cwa triggered interest across a almost {\em all} German districts.

\begin{figure}[t]
  \label{fig:map}
  \vspace{-1em}
  \includegraphics[width=\columnwidth]{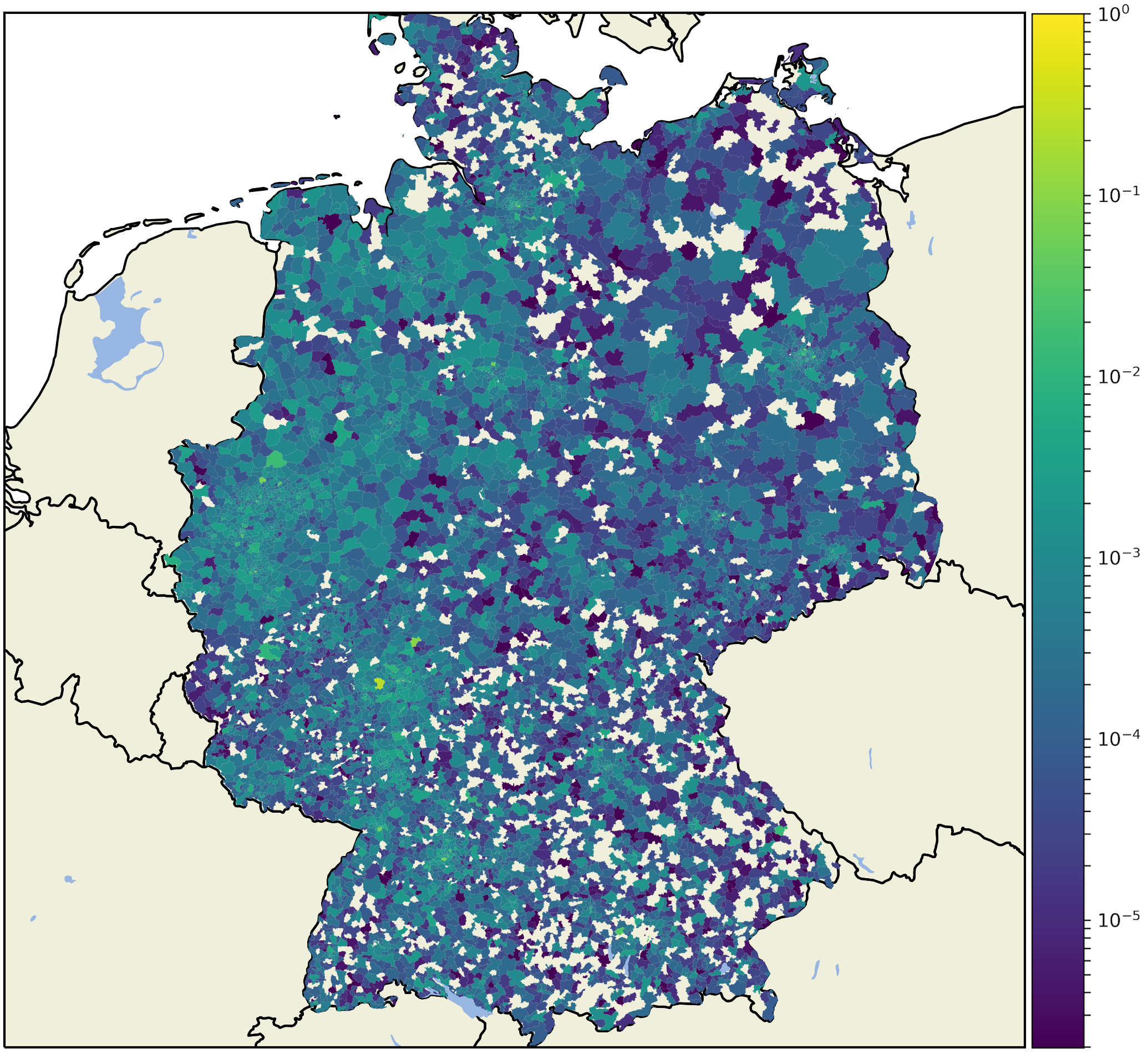}
  \vspace{-2.6em}
  \caption{\cwa traffic by district: usage across Germany aggregated over 10 days normalized by maximum}
  \label{fig:map}
  \vspace{-1.2em}
\end{figure}

\afblock{No effect of local COVID-19 outbreaks.}
Our measurement period contains two local COVID-19 outbreaks: \emph{i)} in Berlin on June 18~\cite{DWNeukoeln}, and \emph{ii)} in G\"utersloh and Warendorf on June 23\cite{DWGuetersloh}.
The latter (June 23) led to few domestic travel restrictions for visitors from these districts~\cite{DWRestrictions}.
While we observe an increase in usage starting on June 23 (see Figure~\ref{fig:overall}), this traffic increase also occurs on federal state level simultaneously---not only in the federal state (NRW) being home to the affected districts.
In G\"utersloh, the traffic increased only very slightly and hardly noticeable (insufficient data for Warendorf).
The outbreak in Berlin on June 18 is only visible for users of a single ISP and not in the overall traffic from Berlin-based users.
For now, we thus conclude that local COVID-19 outbreaks do not appear to generally increase traffic in only the affected regions.
Instead, nation-wide news reports on outbreaks \emph{might} contribute to growing app interest across Germany---an effect we aim to investigate in future work.

\afblock{Conclusion.}
Already on its first day, the \cwa app generated substantial interest---manifested in traffic from almost all German districts.
Local COVID-19 outbreaks do not appear to increase traffic in the affected regions but can correlate to nation-wide increases.
On this basis, we will investigate patterns driving local adoption in future work; e.g., if and how does news media fire \cwa interest; and what will be the long-term app interest.

\bibliographystyle{ACM-Reference-Format}
\bibliography{references}

\end{document}